\documentclass{article}
\usepackage{spconf}
\usepackage{graphicx}
\usepackage{psfrag}
\usepackage{url}
\usepackage{stfloats}
\usepackage{amsmath,epsfig,amsfonts,amssymb,graphics,psfrag,amsthm,calc,subfigure,url,bm,cite,color}
\usepackage{array}
\usepackage{caption}
\usepackage{calc}
\usepackage{subfig}
\usepackage{cases}
\usepackage{verbatim}
\usepackage{algorithm,algorithmic}
\usepackage{multirow}
\usepackage{mathrsfs}

\newtheorem{Lemma}{Lemma}
\newtheorem{Prop}{Proposition}

\theoremstyle{remark}

{\begin{list}{}{
    \settowidth{\labelwidth}{\mbox{\textnormal{#1}}}%
    \setlength{\leftmargin}{\labelwidth+\labelsep}}}%
{\end{list}}

\title{Sum Rate Maximization for Multiuser MISO Downlink with Intelligent Reflecting Surface}
\name{Silei Wang$^\star$, Qiang Li$^\star$, Sissi Xiaoxiao Wu$^\dag$ and Jingran Lin$^\star$}

\address {$^\star$School of Info. \& Comm.  Eng., University of Electronic Science \& Technology of China, P.~R.~China \\
$^\dag$College of Electronics and Information Engineering, Shenzhen University, Shenzhen, P.~R.~China
}

\begin{document}
\bibliographystyle{IEEEtran}
\ninept

\newcommand\bcc[2][c]{\ensuremath{\bm{\mathcal{#2}}}}      
\newcommand\bcl[2][c]{\ensuremath{\bm{#2}}}
\newcommand\Ib{\ensuremath{{\bm I}}}
\newcommand\Vb{\ensuremath{{\bm V}}}
\newcommand\vb{\ensuremath{{\bm v}}}
\newcommand\Hb{\ensuremath{{\bm H}}}
\newcommand\ub{\ensuremath{{\bm u}}}
\newcommand\hb{\ensuremath{{\bm h}}}
\newcommand\xb{\ensuremath{{\bm x}}}
\newcommand\gb{\ensuremath{{\bm g}}}
\newcommand\Thetab{\ensuremath{{\bm \Theta}}}
\newcommand\thetab{\ensuremath{{\boldsymbol \theta}}}
\newcommand\Gb{\ensuremath{{\bm G}}}
\newcommand\Pb{\ensuremath{{\bm P}}}
\newcommand\taub{\ensuremath{{\boldsymbol \tau}}}
\newcommand\zb{\ensuremath{{\bm z}}}
\newcommand\st{\ensuremath{{\rm ~s.t.}}}


\def\blue{\color{blue}}
\def\red{\color{red}}
\definecolor{orange}{RGB}{255,107,0}
\def\orange{\color{orange}}
%

\maketitle
\begin{abstract}
The intelligent reflecting surface (IRS) is a planar array with massive reconfigurable passive elements, which can align the reflecting signals at the receivers via controlling the phase shifts at each element independently. Since IRS can be implemented without RF chain, it is seen as a cost-effective solution for boosting the spectral efficiency for the future communication system. This work considers the IRS-aided multiuser multi-input single-output downlink transmission, and our goal is to maximize the sum rate by jointly optimizing the transmit beamformer at the base station and the phase shifts, which can be either continuous or discrete, at the IRS.
This sum rate maximization (SRM) problem is challenging, especially for the discrete-phase case. To tackle it, we first derive a more tractable equivalent formulation of the SRM problem with structured convex constraints and a smooth objective. From there, a custom-derived block-coordinated accelerated projected gradient algorithm is developed. Simulation results demonstrate that the proposed design outperforms state-of-the-art design in the sum rate performance.
\end{abstract} \vspace{-0.0cm}
\begin{keywords}
Intelligent reflecting surface, passive beamforming, accelerated projected gradient
\end{keywords}


\vspace{-0.23cm}
\section{Introduction} \label{sec:introduction}
\vspace{-5pt}
Low-cost, energy efficient and high spectral efficiency  transmission techniques are indispensable for future communication system. To this end, the intelligent reflecting surface (IRS) has  recently been put forward and gained considerable attention. The idea of IRS is to deploy a large planar array  to intentionally create  additional  reflecting paths for the receivers. More specifically, the IRS consists of a large number of passive  elements, each of which is capable of independently reflecting the incident electromagnetic wave with certain  phase shifts. Therefore, by intelligently adjusting the reflecting phase shift of each element, IRS is able  to adapt to wireless channels, so that the reflected signals can be constructively (resp. destructively) aligned at the desired (resp. non-intended) receiver\footnote{A smart controller integrated in the IRS is responsible for coordinating and exchanging information with other network components (e.g., BSs and users) via a dedicated control link and configuring the phase shift at each reflecting element of the IRS.}~\cite{wu2019towards}, \cite{liaskos2018new}.


While currently IRS is still in its infancy stage, there has been a flow of researches on incorporating IRS into the existing communication systems. In particular, the work~\cite{wu2019intelligent} considered the joint active and passive beamforming to minimize the transmit power at the based station (BS). The sum rate maximization problem is studied in~\cite{yu2019miso} and \cite{guo2020weighted}, where the former focused on the single-user case, and the latter on the multiple users' case. Joint symbol-level precoding and IRS design is considered in~\cite{shao2020minimum},\cite{wang2020onebit} to minimize the symbol-error probability of all the users. Besides the conventional communications, the IRS has been used in improving the physical-layer security \cite{hong2020artificial}, \cite{guan2020intelligent}, assisting the energy transfer~\cite{pan2020intelligent,tang2020joint,Mishra2019} and orthogonal frequency division multiplexing (OFDM) system~\cite{yang2020intelligentr}. We should mention that the aforementioned works on IRS are based on the assumption that the reflection elements are implemented with infinite resolution phase shifters, that is, the phase shifts of the reflection elements can vary continuously from 0 to $2\pi$. However, in practice it could be costly to achieve a continuous phase shifter due to the hardware limitations, especially in the case of a large number of reflection elements at the IRS. As such, some recent studies on IRS have taken into account the finite-resolution phase shifters. The work~\cite{wu2020beamforming} considered the transmit power minimization problem for IRS with finite discrete-phase shifts. In~\cite{zhao2020intelligent}, the authors proposed a new two-timescale (TTS) transmission protocol to maximize the average sum rate, when the phase shifts of IRS take discrete values.

In this work, we consider the IRS-aided multiuser multi-input and single-output (MISO) downlink communication system, where the system sum rate is maximized by jointly optimizing the transmit beamformer at the  BS and the (continuous/discrete) phase shifts at the IRS.
This sum rate maximization (SRM) problem is nonconvex, and furthermore a mixed integer nonlinear program (MINP) when the discrete-phase shifts are concerned. To tackle it, we first derive an equivalent formulation of the SRM problem by using the  recent advances in discrete-phase optimization~\cite{shao2018framework}. The advantage of the reformulation is that it has simple structured convex constraints and a smooth objective function. Based on the reformulation, a block-coordinated accelerated projected gradient (APG)  algorithm is proposed. Simulation results demonstrate that the proposed algorithm is able to achieve better sum rate than a state of the art algorithm under different resolutions of the phase shifters.

\vspace{-6pt}

\section{System Model and Problem Statement} \label{sec:model-prob-statement}
\vspace{-6pt}
Consider a multiuser MISO downlink transmission with the aid of an IRS. The BS equipped with $M$ transmit antennas unicasts $K$ independent information to $K$ single-antenna users. The IRS with $N$ reflecting elements is deployed to assist the downlink communications.
Let $s_k$ with ${\mathbb{E}}\{|s_k|^2\}=1$ and  $\bm v_k \in \mathbb{C}^M$ be the data symbol and the associated transmit beamformer for the $k$th user, resp. The transmit signal $\bm x \in \mathbb{C}^M$ at the BS and the reflected signal $\bm z_r \in \mathbb{C}^N$ at the IRS are resp. given by
\begin{equation} \label{eq:tx_sig}
 \bm x = \textstyle \sum_{k\in \cal K} \bm v_k s_k \quad {\rm and}\quad \bm z_r = {\Thetab}^H \bm G \bm x,
\end{equation}
where $\bm G \in \mathbb{C}^{N\times M}$ is the channel from the BS to the IRS; $\Thetab=\sqrt{\eta} \text{Diag}(\thetab)$ denotes the phase-shift matrix with ${\rm Diag}(\bm \theta)$ being a diagonal matrix with diagonal entries $\bm \theta=[\theta_1,\cdots,\theta_N]^T$, where $\eta \in [0,1]$ and $\theta_n=e^{j\phi_n}$ are the  reflection coefficient and the phase at the $n$-th reflection element of the IRS, resp. In this work, we consider the phase shift at each element of the IRS can take both the continuous values in $(0,2 \pi]$ and a finite number of discrete values equally spaced in $(0,2 \pi]$; that is,
\[  \theta_n \in {\cal F}_{\rm CP} \triangleq \{ \theta \in \mathbb{C}~|~ |\theta| =1   \},~\forall~n=1,\ldots, N,   \]
for the case of continuous phase shifts, and
\[   \theta_n \in {\cal F}_{\rm DP} \triangleq \{ 1, \omega, \ldots, \omega^{L-1} \},~\forall~n=1,\ldots, N,   \]
for the case of discrete phase shifts, where $\omega = e^{\mathfrak{j}2\pi/L}$ and $L$ is the number of realizable phase angles.

Let  $\bm h_{d,k}\in \mathbb{C}^M$ / $\bm h_{r,k}\in \mathbb{C}^N$ be the channel from the BS/IRS to the $k$th user. From~\eqref{eq:tx_sig}, the received signal at the $k$th user is  given by 
\begin{align*}
y_{k} & =\hb_{d,k}^H\xb+\hb_{r,k}^H{\Thetab}^H\Gb\xb+n_k\\
& = (\hb_{d,k}^H+\hb_{r,k}^H{\Thetab}^H\Gb) \textstyle \sum_{\ell\in \cal K} \bm v_\ell s_\ell  +n_k \\
& = (\hb_{d,k}+  \bm H_{r,k} \bm \theta)^H \textstyle \sum_{\ell\in \cal K} \bm v_\ell s_\ell  +n_k,
\end{align*}
where $\bm H_{r,k} \triangleq \sqrt{\eta}  \bm G^H {\rm Diag}(\hb_{r,k} ) $ and $n_k\sim \mathcal{CN}(0,\sigma_k^2)$  is additive white Gaussian noise with mean zero and variance $\sigma_k^2$. Accordingly, the received SINR at the user $k$ is
\begin{align}
\texttt{SINR}_{k}=\frac{\left| (\hb_{d,k}+  \bm H_{r,k} \bm \theta)^H\vb_{k}\right|^{2}}{\sigma_k^{2}+\sum\limits_{j\neq k}{\left| (\hb_{d,k}+  \bm H_{r,k} \bm \theta)^H\vb_{j}\right|^{2}}},\quad \forall~k\in \cal K.
\end{align}

Upon the above model, our problem of interest is to jointly design the transmit beamformers $\{ \bm v_k\}_{k\in \cal K}$ at the BS and the phase shift $\bm \theta$ at the IRS, so that the sum rate of all the users is maximized, viz.,
\begin{subequations}  \label{problem:MAX sumrate}
	\begin{align}
	\max_{\thetab, \{\bm v_k\}_k}~& \textstyle \sum_{k\in \cal K}\text{log}(1+\texttt{SINR}_{k})   \label{problem:MAX sumrate a} \\
	\st ~~&\textstyle\sum_{k\in \cal K}\|\vb_{k}\|^2 \leq P_{\text{max}},  \label{problem:MAX sumrate b} \\
~~& \theta_n \in {\cal F},~\forall~n=1,\ldots,N,
 \label{problem:MAX sumrate c}
	\end{align}
\end{subequations}
where $P_{\max}$ is the total transmit power budget at the BS, and ${\cal F}$ represents either $ {\cal F}_{\rm CP}$ or $ {\cal F}_{\rm DP}$.

The difficulty of solving problem~\eqref{problem:MAX sumrate} lies in the tightly coupled variables in the sum rate expression and the nonconvex phase-shift constraints~\eqref{problem:MAX sumrate c}. Furthermore, when the discrete-phase shift is concerned, problem~\eqref{problem:MAX sumrate} is essentially an MINP, which is generally NP-hard. As such, in the ensuring section, we will focus on developing an efficient approximate approach to problem~\eqref{problem:MAX sumrate}. The crux of our approach is a more tractable reformulation of problem~\eqref{problem:MAX sumrate} and the block-coordinated APG method.

\section{A Tractable Approach to Problem (3)}
Let us first derive an equivalent formulation of problem~\eqref{problem:MAX sumrate}, which facilitates the subsequent algorithm design.
\begin{Prop} \label{theorem:1}
There exists some  $\bar{\lambda}>0$ such that for any $\lambda>\bar{\lambda}$, problem~\eqref{eq:main_eqv} is equivalent to problem~\eqref{problem:MAX sumrate}:
\begin{subequations}\label{eq:main_eqv}
	\begin{align}
\min_{\bm \theta, \{u_k,\vb_k, w_k\}_{k \in \cal K}}~~& \phi_\lambda( \bm V, \bm u, \bm w, \bm \theta) \label{eq:main_eqv_a}\\
\st ~~&\sum_{k\in \cal K}\|\vb_{k}\|^2 \leq P_{\max}, \label{eq:main_eqv_b}\\
~~& w_k \geq 0, ~\forall~k\in \cal K, \label{eq:main_eqv_c}\\
~~ & \theta_n \in \widetilde{\cal F}, ~\forall~n=1,\ldots, N, \label{eq:main_eqv_d}
	\end{align}
\end{subequations}
where $\widetilde{\cal F}$ denotes the convex hull of $\cal F$~\footnote{For ${\cal F}_{\rm CP}$, its convex hull is $\widetilde{\cal F}_{\rm CP} = \{ \theta \in \mathbb{C}~|~|\theta|\leq 1 \}$, and for  ${\cal F}_{\rm DP}$, its convex hull is a regular polygon with vertices $\{1,\omega,\ldots,\omega^{L-1}\}$.} and
\begin{align}
\hspace{-8pt}\phi_\lambda( \bm V, \bm \theta, \bm u, \bm w)  &  = \sum\limits_{k\in \cal K}(w_k e_k(u_k,\Vb)-\log(w_k))  - \lambda \| \bm \theta\|^2 \label{eq:objective}\\
 e_k(u_k,\Vb) &  =\sigma_k^2|u_k|^2+ |1-u_k  (\hb_{d,k}+  \bm H_{r,k} \bm \theta)^H \vb_k|^2 \notag \\
&~~ +\sum\limits_{j \neq k}|u_k  (\hb_{d,k}+  \bm H_{r,k} \bm \theta)^H \vb_j|^2 \label{eq:MSE}
\end{align}
\end{Prop}

\noindent{\it Proof.}~See the Appendix. \hfill $\blacksquare$

The main advantage of the reformulation~\eqref{eq:main_eqv} is that all the constraints are convex with a nice geometry structure, particularly for
the set $\widetilde{\cal F}$. However, the tightly coupled nonconvex objective still poses a challenge for problem~\eqref{eq:main_eqv}. In the following, we will take a divide-and-conquer  approach to alternately optimize problem~\eqref{eq:main_eqv} with respect to $\{u_k, \bm v_k, w_k \}_{k\in \cal K}$ and $\bm \theta$. Algorithm~\ref{alg:main} summarizes the main procedure of our approach. Notice that to avoid an ill-posed problem, we start with a small penalty $\lambda$ and gradually increase it for every $J$ iterations (cf. step 5).

\begin{algorithm}
	\caption{A divide-and-conquer approach to problem~\eqref{eq:main_eqv}}\label{alg:main}
	\begin{algorithmic}[1]
		\STATE
		Given $(\bm V^0, \bm u^0, \bm w^0,\bm \theta^0)$, an initial penalty  $\lambda>0$,  integers $J\geq 1$, $c>1$ and $t=0$.		
		\REPEAT
		\STATE
		\begin{equation}\label{eq:subproblem_1}
		\begin{aligned}
	\hspace{-10pt}	(\bm V^{t+1}, \bm u^{t+1}, \bm w^{t+1}) = \arg\min_{\bm V, \bm u, \bm w} &~ \phi_\lambda( \bm V,  \bm u, \bm w,\bm \theta^t)\\
		{\rm s.t.} &~\eqref{eq:main_eqv_b}-\eqref{eq:main_eqv_c}
		\end{aligned}
		\end{equation}
	\STATE
	\begin{equation}\label{eq:subproblem_2}
\hspace{-10pt}	\bm \theta^{t+1} = \arg\min_{\bm \theta \in \widetilde{\cal F}^N} ~\phi_\lambda( \bm V^{t+1}, \bm u^{t+1}, \bm w^{t+1}, \bm \theta)
	\end{equation}
	\STATE
	Update $\lambda=\lambda c$ every $J$ iterations
\STATE
	$t=t+1$
		\UNTIL{some stopping criterion is satisfied}.
	\end{algorithmic}
\end{algorithm}
\vspace{-10pt}

\subsection{Solution to problem~\eqref{eq:subproblem_1} }
With fixed $\bm \theta$, it is easy to see that problem~\eqref{eq:subproblem_1} is exactly the same as the weighted MMSE (WMMSE) problem in~\cite{shi2011iteratively}. Therefore, a block-coordinate descent (BCD) approach can be employed to handle problem~\eqref{eq:subproblem_1}. Specifically, the WMMSE algorithm cyclically optimizes one of the block variables in $(\bm u,\bm w,\bm V)$ with the other two  fixed. It is shown in~\cite{shi2011iteratively} that each block  variable can be updated in  closed form, which is given by
\begin{subequations}
	\begin{align}
	u_k &= \frac{(\hb_{d,k}+  \bm H_{r,k} \bm \theta)^H \vb_k}{\sigma_k^2+ \sum_{j\in \cal K}| (\hb_{d,k}+  \bm H_{r,k} \bm \theta)^H \vb_j|^{2}},~\forall~k\in \cal K,\\
	w_k  & =e_k^{-1}(u_k ,\Vb),~\forall~k\in \cal K, \\
	\vb_k  &=  u_k w_k \bm F(\bm u, \bm w)\hb_k,~\forall~k\in \cal K,
	\end{align}
\end{subequations}
where
\begin{align*}
 & \bm F(\bm u, \bm w)\\
 = & \left( \sum_{j \in \cal K}{ w_j|u_j|^2   (\hb_{d,j}+  \bm H_{r,j} \bm \theta) (\hb_{d,j}+  \bm H_{r,j} \bm \theta)^H}+\mu \Ib \right)^{-1}
\end{align*} and $\mu \in \mathbb{R}_+$ is Lagrangian multiplier associated with the transmit power constraint~\eqref{eq:main_eqv_b}. The $\mu$ can be determined by bisection search such that the complementary condition holds.

\subsection{Solution to problem~\eqref{eq:subproblem_2}}

Problem~\eqref{eq:subproblem_2} can be simplified as the following problem after dropping the terms irrelevant to $\bm \theta$
\begin{equation}\label{eq:subprob2_eqv}
\begin{aligned}
\min_{\bm \theta} & ~  f_\lambda(\bm \theta) \triangleq  \bm \theta^H  \bm A  \bm \theta + 2{\mathfrak R}( \bm b^H \bm \theta) -  \lambda \| \bm \theta \|^2 \\
{\rm s.t.} & ~ \theta_n \in \widetilde{\cal F},~\forall~n,
\end{aligned}
\end{equation}
where
\begin{align*}
\bm A & =  \sum_{k \in \cal K} w_k |u_k|^2 \bm H_{r,k}^H (\sum_{j\in \cal K} \bm v_j \bm v_j^H) \bm H_{r,k}, \\
\bm b &  = \sum_{k\in \cal K} \sum_{j \in \cal K} \bm b_{k,j}, \\
\bm b_{k,j} & = \begin{cases}
(w_k|u_k|^2 \bm v_k^H \bm h_{d,k} -w_k u_k ) \bm H_{r,k}^H \bm v_k, & j=k, \\
w_k |u_k|^2 \bm H_{r,k}\bm v_j \bm v_j^H \bm h_{d,k}, & j\neq k.
\end{cases}
\end{align*}

Problem~\eqref{eq:subprob2_eqv} is generally a nonconvex problem with the objective $f_\lambda(\bm \theta)$ in the form of  difference-of-convex (DC) functions. We tackle it by gradient extrapolated majorization minimization (GEMM)~\cite{shao2018framework}. Specifically, consider a convex surrogate function $G_\lambda(\bm \theta|\bar{\bm \theta})$ of $f_\lambda(\bm \theta)$, which is given by
\[ G_\lambda(\bm \theta|\bar{\bm \theta}) =  \bm \theta^H  \bm A  \bm \theta + 2{\mathfrak R}( \bm b^H \bm \theta) -   \lambda ( \|\bar{ \bm \theta}\|^2 + 2 {\mathfrak R}(\bar{ \bm \theta}^H ( \bm \theta - \bar{\bm \theta})) ) \]
for some $\bar{\bm \theta} \in \widetilde{\cal F}$; i.e., $G_{\lambda}(\bm \theta | \bar{ \bm \theta})$ approximates $f_\lambda(\bm \theta)$ by linearizing the concave term $-\lambda \| \bm \theta\|^2$ at the point $\bar{ \bm \theta}$. It is easy to check that
\begin{equation*}
f_\lambda(\bm \theta) \leq G_\lambda(\bm \theta|\bar{\bm \theta}), ~\forall~\bm \theta, \bar{\bm \theta} \in \widetilde{\cal F}  ~{\rm and}~
f_\lambda(\bar{\bm \theta}) = G_\lambda(\bar{\bm \theta}|\bar{\bm \theta}), ~\forall~\bar{\bm \theta} \in \widetilde{\cal F},
\end{equation*}
i.e., $G_\lambda(\bm \theta|\bar{\bm \theta})$ is a majorant of $f_\lambda(\bm \theta)$.
Given some initial $\bm \theta^0$, we recursively update $\bm \theta$ via
\begin{equation}\label{eq:MM}
\bm \theta^{\ell+1} = \arg\min_{\bm \theta \in \widetilde{\cal F}} ~ G_\lambda(\bm \theta|\bm \theta^{\ell}), ~\ell=0,1,2,\ldots \vspace{-5pt}
\end{equation}
and output $\bm \theta^\ell$ as an approximate solution of problem~\eqref{eq:subprob2_eqv}.

The remaining issue is how to efficiently calculate the MM update in~\eqref{eq:MM}. Herein, we employ the APG method to complete this task. Let $i=0$, $\hat{\bm \theta}^{i-1} =\hat{\bm \theta}^i = \bm \theta^\ell$. The APG method repeatedly performs the following iterations:
\begin{align}\label{eq:APG}
\hat{\bm \theta}^{i+1} &= \Pi_{\widetilde{\cal F}}\big( \bm z^i - \frac{1}{\rho_i} \nabla_{\bm \theta} G_\lambda(\bm z^i|\bm \theta^\ell) \big), ~~i=0,1,2,\ldots
\end{align}
and outputs the final $\hat{\bm \theta}^i$ as $\bm \theta^{\ell+1}$. In~\eqref{eq:APG}, $1/\rho_i>0$ is the step size, which can be determined by backtracking line search; $\bm z^i$ is an extrapolated point  and is given by
\[\bm z^i =  \hat{\bm \theta}^i + \zeta_i (\hat{\bm \theta}^i - \hat{\bm \theta}^{i-1}) \]
with
$ \zeta_i = \frac{\eta_i -1}{\eta_i}, ~\eta_i = \frac{1+\sqrt{1+4 \eta_{i-1}^2}}{2},~ \eta_{-1} =0,$
and $\Pi_{\widetilde{\cal F}}(\bar{ \bm \theta})$ denotes the element-wise projection of $\bar{\bm \theta}$ onto the set $\widetilde{\cal F}$.
In particular, for $\widetilde{\cal F}_{\rm CP}$, $\Pi_{\widetilde{\cal F}_{\rm CP}} (\theta)$ is given by
	\[ \Pi_{\widetilde{\cal F}_{\rm CP}} (\theta) = \left\{\begin{aligned}
		& \theta, ~~~~~~ \text{if} ~~|\theta|\leq 1, \\ & \theta/|\theta|,  ~~\text{otherwise}.
	\end{aligned} \right. \]
For $\widetilde{\cal F}_{\rm DP}$, $\Pi_{\widetilde{\cal F}_{\rm DP}} (\theta)$ also admits a closed form~\cite{shao2018framework}
	\[ \Pi_{\widetilde{\cal F}_{\rm DP}} (\theta) = e^{\mathfrak{j}\frac{2\pi m}{L}} \left( [{\mathfrak R}(\tilde{\theta})]_0^{\cos(\pi/L)} + \mathfrak{j}  [{\mathfrak I}(\tilde{\theta})]_{-\sin(\pi/L)}^{\sin(\pi/L)}   \right), \]
	where $m=\lfloor \frac{\angle \theta+\pi/L}{2\pi/L} \rfloor$, $\tilde{\theta}=\theta e^{-\mathfrak{j}\frac{2\pi m}{L}}$ and $[\cdot]_a^b$ defines the thresholding operator, i.e., $[x]_a^b= \min\{ b, \max\{x,a\}\}$.

One may notice that for each MM update in~\eqref{eq:MM}, we need to run multiple APG iterations, which could incur high computation burden. A more straightforward and efficient strategy is to perform the MM update  {\it inexactly} by running the APG with only one iteration. By doing so, we can update $\bm \theta^\ell$ more frequently and speed up the convergence. In addition, under some technical conditions~\cite{shao2018framework}, this inexact MM  has been shown to converge to the Karush-Kuhn-Tucker (KKT) solution of problem~\eqref{eq:subprob2_eqv}. Algorithm~\ref{alg:inexact_MM} summarizes the whole procedure for solving  problem~\eqref{eq:subprob2_eqv}.

\begin{algorithm}
	\caption{GEMM Method for Problem \eqref{eq:subprob2_eqv}}\label{alg:inexact_MM}
	\begin{algorithmic}[1]
		\STATE
		{\bf given} $\thetab^0 \in \widetilde{\cal F}^{N\times 1}$, $\epsilon>0$, $\bm \theta^{-1}=\bm \theta^0$ and $\ell=0$.
		\REPEAT
		\STATE
		\begin{equation*}
		\begin{aligned}
		\zb^\ell & = \bm \theta^\ell + \zeta_\ell (\bm \theta^\ell - \bm \theta^{\ell-1}) \\
		\bm \theta^{\ell+1} & = \Pi_{\widetilde{\cal F}}\left( \bm z^\ell - \frac{1}{\rho_\ell} \nabla_{\bm \theta} G_\lambda(\bm z^\ell|\bm \theta^\ell) \right) \\
		\ell&=\ell+1
		\end{aligned}
		\end{equation*}
		\UNTIL{$\|\thetab^{\ell+1}-\thetab^\ell\|<\epsilon$}.
	\end{algorithmic}
\end{algorithm}
\vspace{-10pt}

\section{Simulation Results}
The simulation scenario is as follows: The BS and the IRS are  located at (0,\,0) and (50,\,0), resp., and there are eight users ($K=8$) which are randomly and uniformly distributed within a circle centered at (40,\,20) with radius $10$\,m. We set $M=8$, $\eta=1$ , $\epsilon=10^{-5}$ and $\sigma_k^2=-80\text{dBm},\forall k$. The penalty in Algorithm 1 is initialized as $\lambda=0.01$ and increased by $c=5$ every $J=5$ iterations. The large-scale path loss is modeled as $L(d) = C_0 \zeta d^{-\alpha}$, where $C_0$ is the path loss at the reference distance $1$ meter, $\zeta$ denotes the product of the source and the terminal gain,  $d$ and $\alpha$ denote the link distance and the
path loss exponent, resp. Specifically, for the direct channel between the BS and the user $k$, we set $C_0\zeta_{B,k}=-30\text{dB}, \alpha_{B,k}=3.6,\forall k\in {\cal K}$. Regarding the  IRS-aided link, the concatenate path loss is set as $C_0^2\zeta_{B,I}\zeta_{I,k} (d_{B,I})^{-\alpha_{B,I}} (d_{I,k})^{-\alpha_{I,k}}$ with $C_0^2\zeta_{B,I}\zeta_{I,k}=-40\text{dB}$ and $\alpha_{B,I}=\alpha_{I,k}=2.2$. The small-scale fading follows Rayleigh distribution.
All the results were averaged over 300 channel realizations.


We compare our proposed design with the following schemes: The alternating optimization based algorithm (named ``AO'') in~\cite[Section III]{guo2020weighted}, the block coordinate descent based algorithm (named ``BCD'') in~\cite[Algorithm 2]{guo2020weighted}, which are state-of-the-art algorithms for the IRS-aided SRM problem under the continuous phase shifts. The discrete-phase  counterparts of ``AO''and ``BCD'' are also compared by projecting the continuous phase shifts onto the discrete-phase set via quantization (quant.). As a benchmark, the result of WMMSE without using the IRS~\cite{shi2011iteratively} is also included. The ``AO'' and ``BCD'' algorithms are implemented with the code\footnote{The code is available at https://github.com/guohuayan/WSR-maximization-for-RIS-system} provided in~\cite{guo2020weighted}. The stopping criterion is set as $\| \bm V^t- \bm V^{t-1} \|^2 + \| \bm \theta^t- \bm \theta^{t-1} \|^2 < 10^{-4}$ or the number of outer loop iterations exceeding 30.

\begin{figure}[!h]
 	\centerline{\resizebox{.4\textwidth}{!}{\includegraphics{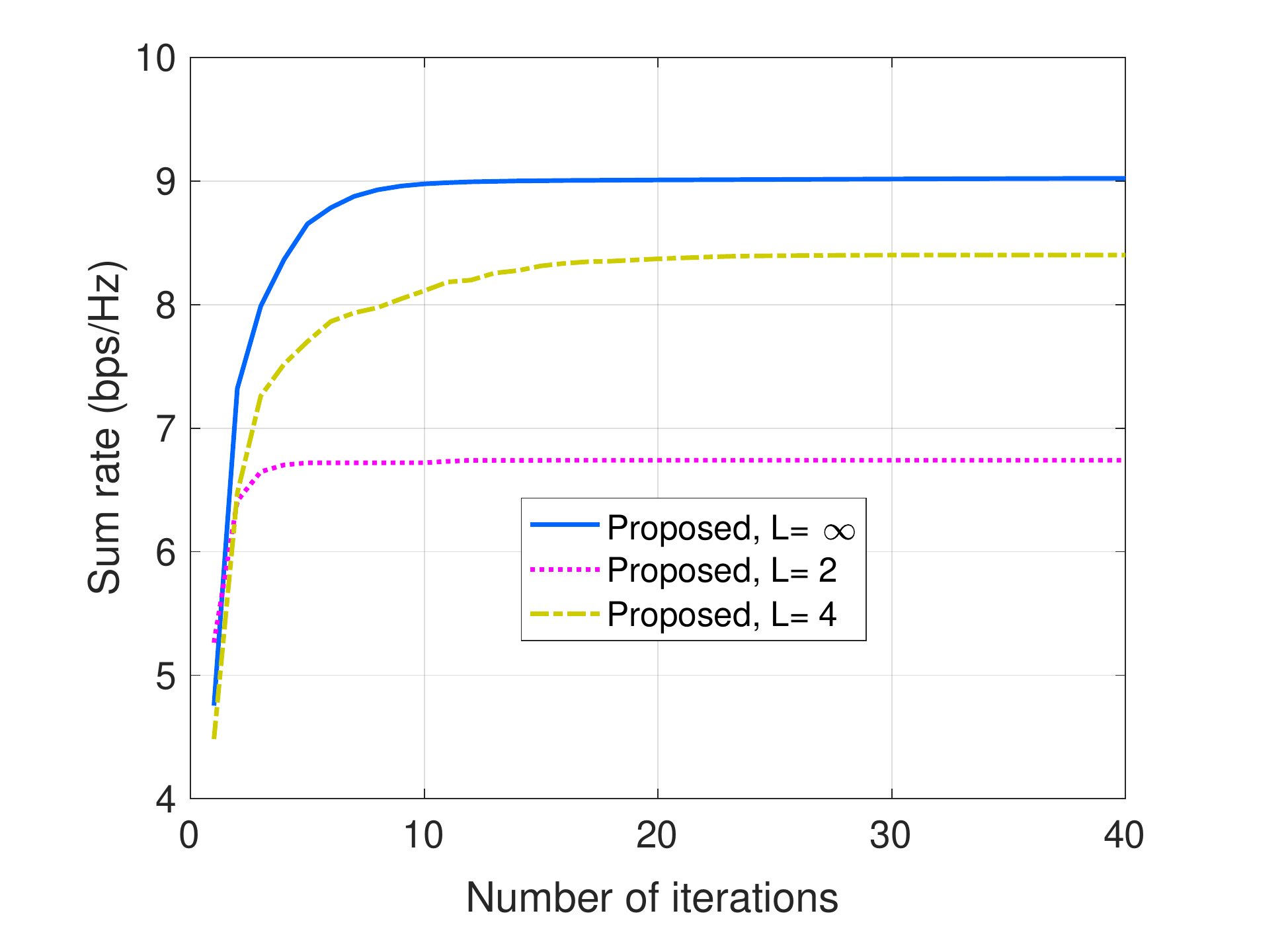}}
 	}\vspace{-10pt}   \caption{Convergence behavior ($P_{\max}=5\text{dBm}$, $N=100$).}
 	\label{fig:convergence}
 	\vspace*{-1.4\baselineskip}
 \end{figure}
\begin{figure}[!h]
	\centerline{\resizebox{.4\textwidth}{!}{\includegraphics{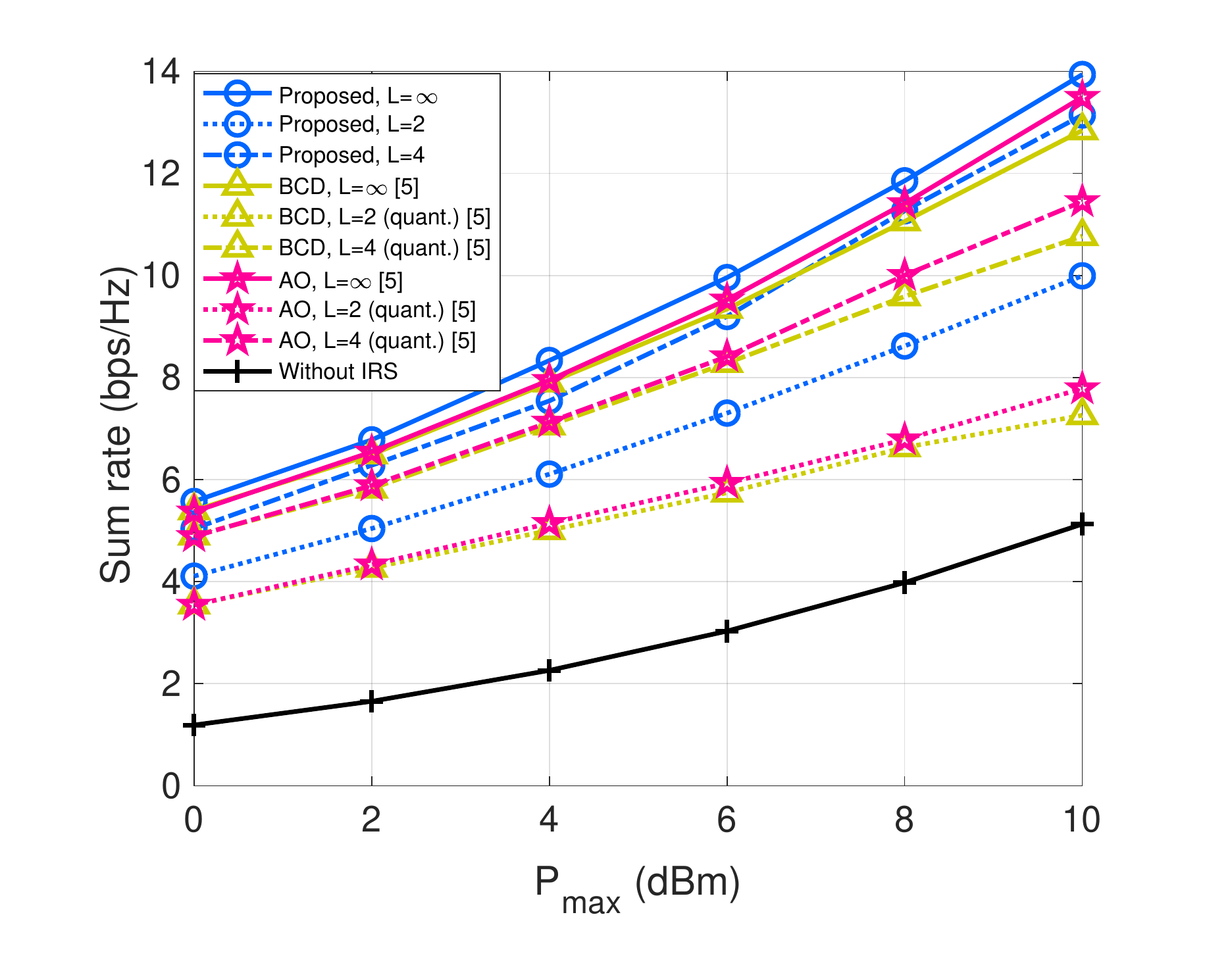}}
	} \vspace{-15pt}  \caption{Sum rate versus $P_{\max}~ (N=100)$.}
	\label{fig:sum_rate}
	\vspace*{-1.4\baselineskip}
\end{figure}
\begin{figure}[!h]
	\centerline{\resizebox{.4\textwidth}{!}{\includegraphics{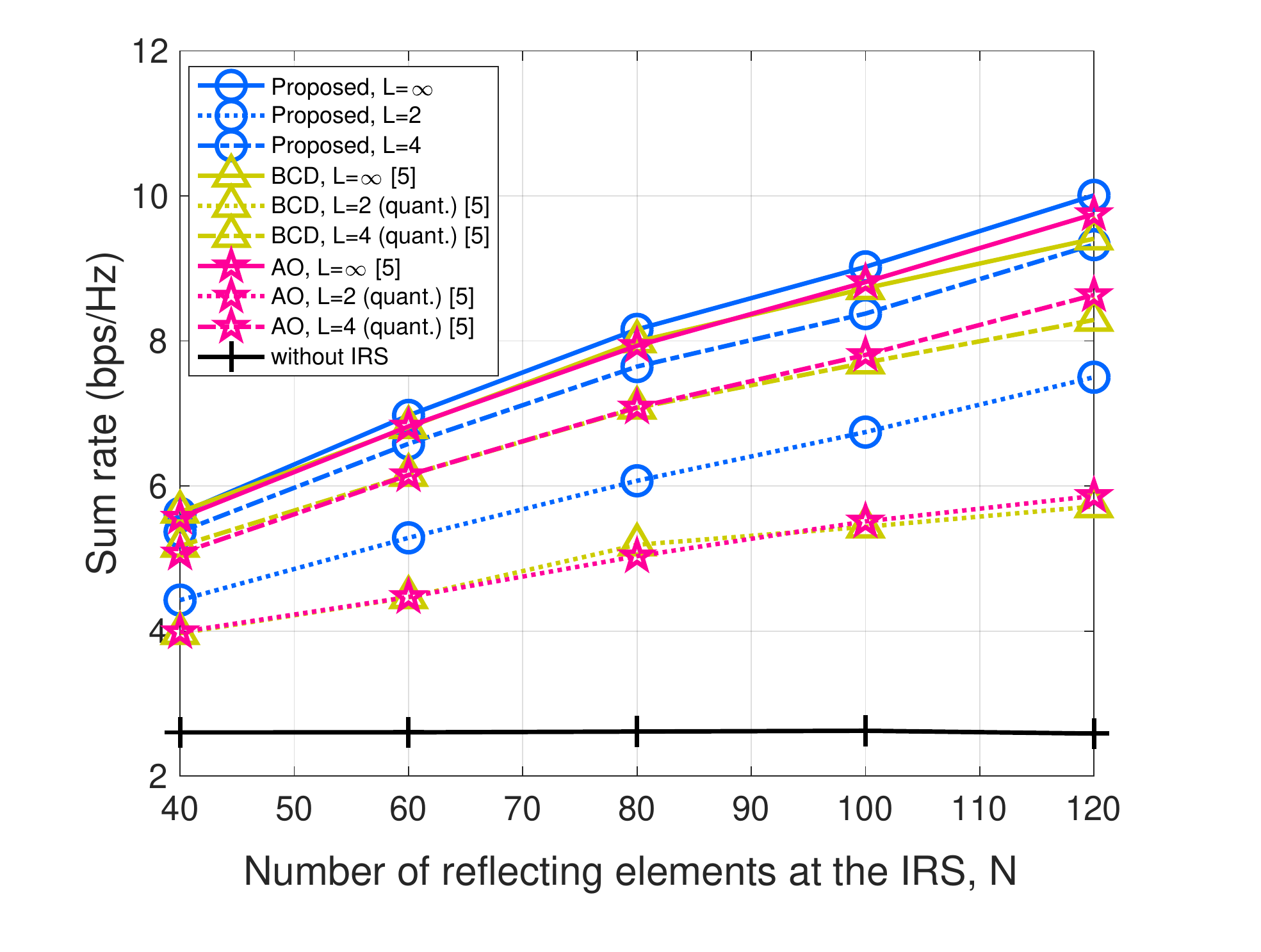}}
	}\vspace{-15pt}   \caption{The sum rate versus $N$, ($P_{\max}=5$dBm).}
	\label{fig:sum_rate_N}
	\vspace*{-1.4\baselineskip}
\end{figure}

Fig.~\ref{fig:convergence} shows the convergence of Algorithm 1 for $N=100$ and $P_{\max}=5\text{dBm}$. It is seen that the proposed algorithm converges quickly within a few iterations under different resolutions of the phase shifters. Fig.~\ref{fig:sum_rate} illustrates the sum rate performance of different designs under different transmit powers with $N=100$. For the continuous-phase shift case (i.e., $\mathcal{F}=\mathcal{F}_{\rm CP}$), one can see that our proposed design achieves better sum rate performance than the algorithms in~\cite{guo2020weighted}. For the discrete-phase shift case (i.e., $\mathcal{F}=\mathcal{F}_{\rm DP}$), ``Proposed, $L=2$'' outperforms ``AO, $L=2$ (quant.)'' and ``BCD, $L=2$ (quant.)'' when 1-bit IRS is considered.
Similar results can be observed for the case of 2-bit IRS ($L=4$). This indicates that it is crucial to explicitly take into account the discrete-phase characteristics of the IRS during the optimization process, and naive quantization could incur serious performance degradation. Moreover, all the IRS-aided schemes attain much higher sum rate than that without IRS. This verifies the effectiveness of the IRS in boosting the system capacity. Fig.~\ref{fig:sum_rate_N} compares the sum rate of different schemes with different number of reflecting elements at the IRS for $P_\text{max}=5\text{dBm}$. Again, the proposed design outperforms the other schemes for all the tested number of IRS elements.

\vspace{-10pt}

\section{Conclusion}
In this paper, we have investigated the joint transmit beamforming and reflecting phase shift design for the IRS-aided MU-MISO downlink transmission. The sum rate maximization problem under both the continuous-phase and the discrete-phase constraints on the IRS is studied. By leveraging on the recent advances in discrete-phase optimization, a block-coordinated APG algorithm is custom-derived.  Numerical results have demonstrated that the inclusion of IRS can effectively improve the spectral efficiency for the conventional wireless systems. Moreover, the proposed design can attain higher sum rate than state-of-the-art designs, especially for the case of discrete-phase shifts.

\vspace{-10pt}

\section{Appendix}\label{sec:appendix}
Firstly, by employing the well-known rate-MMSE relationship~\cite{shi2011iteratively}, we have
\begin{equation}\label{eq:rate-mmse}
\text{log}(1+\texttt{SINR}_{k}) = \log([e_k^{\texttt{MMSE}}]^{-1}),
\end{equation}
where $e_k^{\texttt{MMSE}}$ represents the MMSE of user $k$'s received signal $s_k$, and it is given by
\begin{equation}\label{eq:MMSE}
e_k^{\texttt{MMSE}} = \min_{u_k \in \mathbb{C}}~e_k(u_k, \bm V),
\end{equation}
where $e_k(u_k, \bm V)$ is defined in~\eqref{eq:MSE}. It follows from Eqns.~\eqref{eq:rate-mmse} and \eqref{eq:MMSE} that
\begin{equation}\label{eq:rate-mmse-eqv}
\begin{aligned}
\text{log}(1+\texttt{SINR}_{k}) & =\log([\min_{u_k \in \mathbb{C}}~e_k(u_k, \bm V)]^{-1}) \\
 & = \max_{u_k\in \mathbb{C}} \log([e_k(u_k, \bm V)]^{-1}).
\end{aligned}
\end{equation}
Notice the following identity
\vspace{-3pt}
\[ \log(x^{-1}) = \max_{w \geq 0} -wx+ \log(w) + 1 \]
with the optimal $w^\star = x^{-1}$. The last line of~\eqref{eq:rate-mmse-eqv} can be further expressed as
\begin{equation}\label{eq:rate-mmse-eqv2}
\text{log}(1+\texttt{SINR}_{k})  = \max_{u_k\in \mathbb{C}, w_k \geq 0} ~ -w_k e_k(u_k, \bm V) + \log(w_k) + 1.
\end{equation}
\vspace{-5pt}
Therefore, problem~\eqref{problem:MAX sumrate} is equivalent to
\begin{subequations}\label{eq:rate-mmse-eqv3}
\begin{align}
\min_{\thetab,\bm V , u_k\in \mathbb{C}, w_k \geq 0}~&\sum_{k\in \cal K}  w_k e_k(u_k, \bm V) -\log(w_k)   \label{eq:rate-mmse-eqv3-a} \\
\st ~~&\sum_{k\in \cal K}\|\vb_{k}\|^2 \leq P_{\text{max}},  \label{eq:rate-mmse-eqv3-b} \\
~~& \theta_n \in {\cal F},~\forall~n=1,\ldots,N. \label{eq:rate-mmse-eqv3-c}
\end{align}
\end{subequations}

Next, we leverage the following lemma to tackle the discrete-phase constraints.
\begin{Lemma}[\cite{shao2018framework}] \label{lemma:a negative penalty method}
	Consider the following two  problems
	\begin{equation}  \label{problem:a discrete phase problem}
	\begin{aligned}
	\min_{\xb \in \mathbb{C}^N} &~ f(\bm x) \quad {\rm s.t.}~x_n\in {\cal F}, ~\forall~n=1,\ldots, N,
	\end{aligned}
	\end{equation}
	and
		\begin{equation} \label{problem:a discrete phase problem equivalent}
	\begin{aligned}
	\min_{\xb \in \mathbb{C}^N} &~ f(\bm x) - \lambda \| \bm x \|^2~~ {\rm s.t.}~x_n\in \widetilde{\cal F}, ~\forall~n=1,\ldots N,
	\end{aligned}
	\end{equation}
	where $\lambda>0$, $f: \mathbb{C}^N \to \mathbb{R}$ is a $\mu$-Lipschitz continuous function, and $\widetilde{\cal F}$ denotes the convex hull of $\mathcal{F}$. Then, there exists a constant $\bar{\lambda}>0$ such that for any $\lambda > \bar{\lambda}$, any (globally)
	optimal solution to problem~\eqref{problem:a discrete phase problem} is also a (globally) optimal
	solution to problem~\eqref{problem:a discrete phase problem equivalent}; the converse is also true. Specifically, we have $\bar{\lambda} = \mu$ for the continuous-phase ${\cal F}_{\rm CP}$ case,  and $\bar{\lambda} =  \mu/\sin(\pi/L)$ for the discrete-phase ${\cal F}_{\rm DP}$ case.
\end{Lemma}
By applying Lemma~\ref{lemma:a negative penalty method} for problem~\eqref{eq:rate-mmse-eqv3}, we arrive at the result in Proposition~\ref{theorem:1}. This completes the proof.

\bibliography{IRS_bib}

\end{document}